\documentstyle[11pt]{article}

\newcommand{\sect}[1]{\setcounter{equation}{0}\section{#1}}

\begin{document}
\begin{titlepage}
\hfill{UQMATH-93-10}
\hfill{hep-th/9310183}
\vskip.3in
\begin{center}
{\huge Infinite Families of Gauge-Equivalent $R$-Matrices
and Gradations of Quantized Affine Algebras}
\vskip.3in
{\Large Anthony J. Bracken$~^a$, Gustav W. Delius$~^b$, Mark D. Gould$~^a$}\\
and\\ {\Large Yao-Zhong Zhang$~^a$}
\vskip.3in
$a$. {\large Department of Mathematics, University of Queensland, Brisbane
Qld 4072, Australia}\\
$b$. {\large Fakult\"at f\"ur Physik, Universit\"at Bielefeld,
33501 Bielefeld, Germany}
\end{center}
\vskip.6in
\begin{center}
{\bf Abstract:}
\end{center}
Associated with the fundamental
representation of a quantum algebra such as
$U_q(A_1)$ or $U_q(A_2)$,  there
exist infinitely many gauge-equivalent $R$-matrices with different
spectral-parameter dependences. It is shown how these can be obtained by
examining the infinitely many possible gradations of the corresponding
quantum affine algebras, such as $U_q(A_1^{(1)})$ and $U_q(A_2^{(1)})$, and
explicit formulae are obtained for those two cases. Spectral-dependent
similarity (gauge) transformations relate the $R$-matrices
in different gradations. Nevertheless, the choice of gradation can be
physically significant, as is illustrated in the case of
quantum affine Toda field theories.

\vskip 6cm
\end{titlepage}

\sect{Introduction}

Quantized universal enveloping algebras (quantum algebras)
\cite{Drinfeld}\cite{Jimbo}\cite{Reshetikhin} provide a powerful tool
for solving the spectral-dependent quantum Yang-Baxter equation (QYBE)
\cite{Jones}\cite{ZGB}\cite{Ma et al}\cite{ZG2}, which plays a central
role in the study of integrable systems in many areas of physics.
In particular, each solution is a spectral-dependent $R$-matrix
which may define the Boltzmann weights of
a solvable vertex model in statistical mechanics, or a scattering
matrix in a quantum affine Toda theory.

In this paper,
we present a method for constructing  different spectral-dependent
and gauge equivalent $R$-matrices
associated with one and the same representation of a quantum algebra.
The idea is to examine various gradations of a corresponding quantum affine
algebra, including in particular the important homogeneous and principal
gradations \cite{Belavin}\cite{Jimbo}.
We exploit techniques previously developed
\cite{KT}\cite{ZG2}, which relate spectral-dependent $R$-matrices
associated with a quantum algebra (such as $U_q(A_n)$)
to the universal $R$-matrix of a
corresponding quantum affine algebra (such as $U_q(A_n^{(1)}$).
Here we consider $n = 1,2$, and construct infinitely many gauge equivalent
$R$-matrices with different spectral-dependences, corresponding to
the infinitely many different gradations of each of the
quantum affine algebras $U_q(A_1^{(1)})$ and
$U_q(A_2^{(1)})$.

Gauge equivalent $R$-matrices are known to lead to solvable statistical
models which are essentially equivalent \cite{BP}. In the case of
quantum affine Toda theories, the choice of gradation is more significant,
as we shall show.

\sect{Universal $R$-Matrix for $U_q(A_1^{(1)})$ and $U_q(A_2^{(1)})$}
This section is devoted to a brief review of the
construction of the universal $R$-matrix for $U_q(A_1^{(1)})$ \cite{KT}
and for $U_q(A_2^{(1)})$ \cite{KT}\cite{ZG}.
Throughout the paper, we use the notations:
\begin{eqnarray}
&&({\rm ad}_qx_\alpha)x_\beta=[x_\alpha\,,\,x_\beta]=x_\alpha x_\beta
  -q^{(\alpha,\beta)}x_\beta x_\alpha\nonumber\\
&&\theta(q^h)=q^{-h}\,,~~\theta(E_i)=F_i\,,~~\theta(F_i)=E_i
  \,,~~\theta(q)=q^{-1}\nonumber\\
&&(n)_q=\frac{1-q^{-n}}{1-q^{-1}}\,,~~[n]_q=\frac{q^n-q^{-n}}{q-q^{-1}}\,,~~
  q_\alpha=q^{(\alpha,\alpha)}\nonumber\\
&&{\rm exp}_q(x)=\sum_{n\geq 0}\frac{x^n}{(n)_q!}\,,~~(n)_q!=
  (n)_q(n-1)_q\,...\,(1)_q
\,.\end{eqnarray}

We start with the rank 2 case, and fix
the normal ordering in the positive root system $\Delta_+$ of $A_1^{(1)}$ as
\begin{equation}
\alpha,\, \alpha+\delta,\, \cdots,\, \alpha+n\delta,\, \cdots,\,
\delta,\, 2\delta,\, \cdots,\,
m\delta,\, \cdots\,,\, \cdots\,,\, (\delta-\alpha)+l\delta,\, \cdots
\,, \, \delta-\alpha\,,
\end{equation}
where $\alpha\,,~\delta-\alpha$ are simple roots and $\delta$ is the
minimal positive imaginary root. Then one finds \cite{KT}
the universal $R$-matrix for $U_q(A_1^{(1)})$,
\begin{eqnarray}
R&=&\left (\prod_{n\geq 0}\;{\rm exp}_{q_\alpha}((q-q^{-1})(E_{\alpha+n\delta}
  \otimes  F_{\alpha+n\delta}))\right )\nonumber\\
& &\cdot{\rm exp}\left ( \sum_{n>0}n[n]_{q_\alpha}^{-1}
  (q_\alpha-q_\alpha^{-1})(E_{n\delta}\otimes F_{n\delta})\right )\nonumber\\
& &\cdot\left (\prod_{n\geq 0}\;{\rm exp}_{q_\alpha}((q-q^{-1})
  (E_{(\delta-\alpha)+n\delta}\otimes
  F_{(\delta-\alpha)+n\delta}))\right )\cdot
  q^{\frac{1}{2}h_\alpha\otimes h_\alpha+c\otimes d+d\otimes c}\label
  {sl2R}\,,
\end{eqnarray}
where $c=h_\alpha+h_{\delta-\alpha}$ and the Cartan-Weyl generators,
$E_\gamma\,,~~F_\gamma=\theta(E_\gamma)\,,~~\gamma\in\Delta_+$\,,
are given by
\begin{eqnarray}
&&\tilde{E_\delta}=[(\alpha,\alpha)]_q^{-1}[E_\alpha,\,E_{\delta-
  \alpha}]_q\,,~~~~
  E_{\alpha+n\delta}=(-1)^n\left ({\rm ad}_q\tilde{E_\delta}\right )^nE_\alpha
  \nonumber\\
&&E_{(\delta-\alpha)+n\delta}=\left ({\rm ad}_q\tilde{E_\delta}\right )^n
  E_{\delta-\alpha}\,,~~~...\,,~~~
  \tilde{E}_{n\delta}= [(\alpha,\alpha)]_q^{-1}[E_{\alpha+(n-1)\delta},
  \,E_{\delta-\alpha}]_q\nonumber\\
&&\tilde{E}_{n\delta}=\sum_{p_1+2p_2+\cdots+np_n=n}
  \frac{\left ( q^{(\alpha,\alpha)}-q^{-(\alpha,\alpha)}\right )^{\sum_ip_i-1}}
  {p_1!\;...\;p_n!}(E_{\delta})^{p_1}(E_{2\delta})^{p_2}...
  (E_{n\delta})^{p_n}\,.
\end{eqnarray}
The order in the product (\ref{sl2R}) coincides
with the chosen normal order.

Turning to the rank 3 case, we fix a normal order in the positive root system
$\Delta_+$ of $A_2^{(1)}$\,as
\begin{eqnarray}
&&\alpha,\,\alpha+\beta,\,\alpha+\delta,\,\alpha+\beta+\delta,\,
   ...\,,...\,,\,\alpha+m_1\delta,\,
  \alpha+\beta+m_2\delta,\,...\,,\cdots\,,\nonumber\\
&&\beta,\,\beta+\delta,\,...\,,\,\beta+m_3\delta,\,...\,,\,\delta,\,
  2\delta,\,...\,,\,
  k\delta,\,...\,,\,...\,,\,(\delta-\beta)+l_1\delta,\,...\,,\,
  \delta-\beta,\,...\,,\,\nonumber\\
&&...\,,(\delta-\alpha)+l_2\delta,\,(\delta-\alpha-\beta)+l_3\delta,\,
  ...\,,\,...\,,\,\delta-\alpha,\,\delta-\alpha-\beta\label{order2}\,,
\end{eqnarray}
where $m_i,k,l_i \geq 0\,,~~i=1,2,3$. Then one can show \cite{KT} (see also
\cite{ZG}) that the universal $R$-matrix for $U_q(A_2^{(1)})$ is given by
\begin{eqnarray}
R&=&\prod_{\gamma<\delta}~{\rm exp}_{q_\gamma}
   \left ((q-q^{-1})(E_\gamma\otimes F_\gamma)\right )\nonumber\\
& &\cdot {\rm exp}\left (\sum_{n>0}\sum^2_{i,j=1}
  C^n_{ij}(q)(q-q^{-1})(E^{(i)}_{n\delta}\otimes F^{(j)}_{n\delta})
  \right )\nonumber\\
& &\cdot \prod_{\gamma>\delta}~{\rm exp}_{q_{\gamma}}
  \left ((q-q^{-1})(E_\gamma\otimes F_\gamma)\right )\nonumber\\
& &\cdot q^{\sum^2_{i,j=1}\,(a^{-1}_{\rm sym})^{ij}h_i\otimes h_j
  +c\otimes d+d\otimes c}\label{aR}\,,
\end{eqnarray}
where $c=h_0+h_\psi$ with $\psi=\alpha+\beta$ being
the highest root of $A_2^{(1)}$ and
$(a_{\rm sym}^{ij})=\left (
\begin{array}{cc}
2 & -1\\
-1 & 2
\end{array} \right )$~;~
\begin{equation}
(C^n_{ij}(q))=(C^n_{ji}(q))=
 \frac{n}{[n]_q}\,\frac{[2]_q^2}{q^{2n}+1+q^{-2n}}\,\left (
\begin{array}{cc}
q^n+q^{-n} & (-1)^n\\
(-1)^n & q^n+q^{-n}
\end{array} \right )\,;
\end{equation}
and  the Cartan-Weyl generators, $E_\gamma\,,~~F_\gamma=\theta(E_\gamma)
\,,~~\gamma\in\Delta_+$, are given by
($\alpha_i=\alpha,\,\beta,\,\alpha+\beta$ below)
\begin{eqnarray}
&&E_{\alpha+\beta}=[E_\alpha\,,\,E_\beta]_q\,,~~~~
  E_{\delta-\alpha}=[E_\beta\,,\,E_{\delta-\alpha-\beta}]_q\,,~~~~
  E_{\delta-\beta}=[E_\alpha\,,\,E_{\delta-\alpha-\beta}]_q\nonumber\\
&&\tilde{E}_\delta^{(i)}=[(\alpha_i,\alpha_i)]_q^{-1}[E_{\alpha_i},\,
  E_{\delta-\alpha_i}]_q\,,~~~
  E_{\alpha_i+n\delta}=(-1)^n\left ({\rm ad}_q\tilde{E}_\delta^{(i)}\right )^n
  E_{\alpha_i}\nonumber\\
&&E_{\delta-\alpha_i+n\delta}=\left ({\rm ad}_q\tilde{E}_\delta^{(i)}\right )^n
  E_{\delta-\alpha_i}\,,~~~  ...\,,~~~
  \tilde{E}_{n\delta}^{(i)}= [(\alpha_i,\alpha_i)]_q^{-1}[E_{\alpha_i
  +(n-1)\delta},\,E_{\delta-\alpha_i}]_q \nonumber\\
&&\tilde{E}^{(i)}_{n\delta}=\sum_{p_1+2p_2+\cdots+np_n=n}
  \frac{\left ( q^{(\alpha_i,\alpha_i)}-q^{-(\alpha_i,\alpha_i)}
  \right )^{\sum_ip_i-1}}
  {p_1!\;...\;p_n!}(E^{(i)}_{\delta})^{p_1}(E^{(i)}_{2\delta})^{p_2}...
  (E^{(i)}_{n\delta})^{p_n}\,.
\end{eqnarray}
Once again, the order in the product (\ref{aR}) is defined by that in
(\ref{order2}).

\sect{Infinitely Many Gauge-Equivalent $R$-Matrices for $U_q(A_1)$}
It can be shown that, for any $z\in {\bf C}^\times$,  there exist algebra
homomorphisms
${\rm ev}_z $ :
$U_q(A^{(1)}_1)\rightarrow U_q(A_1)$ given by
\begin{eqnarray}
&&{\rm ev}_z (E_\alpha)=z^{s_1}E_\alpha\,,~~{\rm ev}_z (F_\alpha)=z^{-s_1}
  F_\alpha\,,~~
  {\rm ev}_z (h_\alpha)=h_\alpha\,,~~{\rm ev}_z (c)=0\nonumber\\
&&{\rm ev}_z (E_{\delta-\alpha})=z^{s_0}F_\alpha\,,~~
{\rm ev}_z
  (F_{\delta-\alpha})
  =z^{-s_0}E_\alpha\,,~~
  {\rm ev}_z (h_{\delta-\alpha})=-h_\alpha\,,\label{g1}
\end{eqnarray}
Each homomorphism ${\rm ev}_z $ defines
a corresponding gradation $(s_0,s_1)$ of $U_q(A_1^{(1)})$. $s_0$ and $s_1$
are arbitrary real numbers.

Following similar lines to those developed earlier \cite{ZG2}, we derive
from (\ref{sl2R})
the spectral-dependent universal $R$-matrix
for $U_q(A_1)$,~
corresponding to the gradation $(s_0,s_1)$:
\begin{eqnarray}
R^{\rm (s_0,s_1)}(u)&=&\prod_{n\geq 0}\,{\rm exp}_{q_\alpha}
  \left ( (q-q^{-1})
  u^{(s_1+s_0)n+s_1}
  \left (q^{-nh_\alpha}E_\alpha\otimes F_\alpha
  q^{nh_\alpha}\right )\right )\nonumber\\
& &\cdot{\rm exp}\left ( \sum_{n>0}n[n]_{q_\alpha}^{-1}
  (q_\alpha-q_\alpha^{-1})u^{(s_1+s_0)n}
  (E'_{n\delta}\otimes F'_{n\delta})\right )\nonumber\\
& &\cdot\prod_{n\geq 0}\;{\rm exp}_{q_\alpha}\left ((q-q^{-1})
  u^{(s_1+s_0)n+s_0}
  \left (F_\alpha q^{-nh_\alpha}\otimes q^{nh_\alpha}
  E_\alpha\right )\right )\cdot
  q^{\frac{1}{2}h_\alpha\otimes h_\alpha}\label{loop-sl2R}\,,\label{uni1}
\end{eqnarray}
where $E'_{n\delta}$ and $F'_{n\delta}$ are determined by
the following equalities of formal power series:
\begin{eqnarray}
&&1+(q_\alpha-q_\alpha^{-1})\sum_{k=1}^\infty \tilde{E}'_
  {k\delta}u^k={\rm exp}\left ( (q_\alpha-q_\alpha^{-1})\sum_{l=1}^\infty
  E'_{l\delta}u^l\right )\nonumber\\
&&1-(q_\alpha-q_\alpha^{-1})\sum_{k=1}^\infty \tilde{F}'_
  {k\delta}u^{-k}={\rm exp}\left ( -(q_\alpha-q_\alpha^{-1})\sum_{l=1}^\infty
  F'_{l\delta}u^{-l}\right )\nonumber\\
&&\tilde{E}'_{n\delta}=[2]_q^{-1}(-1)^{n-1}q^{-(n-1)h_\alpha}\left (
  E_\alpha F_\alpha -q^{-2n}F_\alpha E_\alpha\right )\nonumber\\
&&\tilde{F}'_{n\delta}=[2]_q^{-1}(-1)^{n-1}q^{(n-1)h_\alpha}\left (
  F_\alpha E_\alpha -q^{2n}E_\alpha F_\alpha\right )\label{imaginary}\,.
\end{eqnarray}

On applying (\ref{loop-sl2R}) to the concrete representation
on the tensor product space $V_{1/2}\otimes V_{1/2}$, where
$V_{1/2}$ carries the fundamental representation of $U_q(A_1)$, we get
an infinite family of
$R$-matrices with different spectral-dependences,
corresponding to the infinite number of gradations $(s_0,s_1)$,
\begin{equation}
R^{\rm (s_0,s_1)}_{1/2,1/2}(u)=f_q(u)\cdot \left (
\begin{array}{cccc}
1 & {} & {} & {}\\
{} & \frac{q^{-1}(1-u^{s_1+s_0})}{1-q^{-2}u^{s_1+s_0}} &
     \frac{u^{s_1}(1-q^{-2})}{1-q^{-2}u^{s_1+s_0}} & {}\\
{} & \frac{u^{s_0}(1-q^{-2})}{1-q^{-2}u^{s_1+s_0}} &
     \frac{q^{-1}(1-u^{s_1+s_0})}{1-q^{-2}u^{s_1+s_0}} & {}\\
{} & {} & {} & 1
\end{array}
\right )\,,\label{1/2,1/2}
\end{equation}
where $f_q(u)$ is an irrelevant overall scalar factor,
\begin{equation}
f_q(u)=q^{1/2}\cdot {\rm exp}\left (\sum_{n>0}\frac{q^n-q^{-n}}{q^n+q^{-n}}
\frac{u^{(s_1+s_0)n}}{n}\right )
\,,\end{equation}
which will be ignored in what follows. It is readily checked directly that
each of the $R$-matrices (12) satisfies the parameter-dependent QYBE.

Some remarks are in order:\\
(i) Semi-classical limit:
\begin{eqnarray}
R^{(s_0,s_1)}_{1/2,1/2}(u)&=&I+\left (\frac{1}{2}\log q\right )
  r^{(s_0,s_1)}_{1/2,1/2}(u)+{\cal O}\left ((\frac{1}{2}\log q)^2
  \right )\,,\nonumber\\
r^{\rm (s_0,s_1)}_{1/2,1/2}(u)&=& \left (
\begin{array}{cccc}
\frac{1+u^{s_1+s_0}}{1-u^{s_1+s_0}} & {} & {} & {}\\
{} & -\frac{1+u^{s_1+s_0}}{1-u^{s_1+s_0}} &
     \frac{4u^{s_1}}{1-u^{s_1+s_0}} & {}\\
{} & \frac{4u^{s_0}}{1-u^{s_1+s_0}} &
     -\frac{1+u^{s_1+s_0}}{1-u^{s_1+s_0}} & {}\\
{} & {} & {} &
\frac{1+u^{s_1+s_0}}{1-u^{s_1+s_0}}
\end{array}
\right )\,.\label{classical1}
\end{eqnarray}
Eq.(14) defines the corresponding rational solutions of the classical
Yang-Baxter equation (CYBE).
\\(ii) Homogeneous gradation $s_1=0\,,~s_0=1$:
This reproduces a result well known in the literature
\cite{Jimbo}; we denote it by
$R^{\rm h}_{1/2,1/2}(u)$:
\begin{equation}
R^{\rm h}_{1/2,1/2}(u)= \left (
\begin{array}{cccc}
1 & {} & {} & {}\\
{} & \frac{q^{-1}(1-u)}{1-q^{-2}u} &
     \frac{1-q^{-2}}{1-q^{-2}u} & {}\\
{} & \frac{u(1-q^{-2})}{1-q^{-2}u} &
     \frac{q^{-1}(1-u)}{1-q^{-2}u} & {}\\
{} & {} & {} & 1
\end{array}
\right )\label{1/2,1/2-h}\,.
\end{equation}
\\(iii) Principal gradation $s_1=s_0=1$: This produces the symmetric form of
$R$-matrix (see, cf. \cite{Wadati}); we denote it by
$R^{\rm p}_{1/2,1/2}(u)$:
\begin{equation}
R^{\rm p}_{1/2,1/2}(u)= \left (
\begin{array}{cccc}
1 & {} & {} & {}\\
{} & \frac{q^{-1}(1-u^2)}{1-q^{-2}u^2} &
     \frac{u(1-q^{-2})}{1-q^{-2}u^2} & {}\\
{} & \frac{u(1-q^{-2})}{1-q^{-2}u^2} &
     \frac{q^{-1}(1-u^2)}{1-q^{-2}u^2} & {}\\
{} & {} & {} & 1
\end{array}
\right )\,.\label{1/2,1/2-p}
\end{equation}
(iv) It is important to note that, with the exception of
$R^{\rm h}_{1/2,1/2}(u)$, none of these $R$-matrices can be {\em directly}
obtained by solving the Jimbo equations \cite{Jimbo} or by using
Yang-Baxterization procedures developed previously
\cite{Jones}\cite{ZGB}\cite{Ma et al}. In particular, it is only for
the homogeneous gradation that $PR^{(s_0,s_1)}_{1/2,1/2}(u)$
commutes with all generators of
$U_q(A_1)$, where $P$ is the operator that permutes the two spaces in the
tensor product $V_{1/2}\otimes V_{1/2}$. \\
(v) The various
$R^{\rm (s_0,s_1)}_{1/2,1/2}(u)$ are related to $R^{\rm h}_{1/2,1/2}(u)$
by spectral-dependent similarity (gauge) transformations,
\begin{eqnarray}
&&S(u)R^{\rm (s_0,s_1)}_{1/2,1/2}(u)S^{-1}(u)=R^{\rm h}_{1/2,1/2}
  (u^{s_1+s_0})\nonumber\\
&&S(u)={\rm diag}\left (1\,,\, u^{-s_1/2}\,,\,u^{s_1/2}\,,\,
  1\right )\label{trans1}\,,
\end{eqnarray}
Note the change in the spectral parameter on the right hand side.
This defines a gauge symmetry of the spectral-dependent QYBE.
Thus the differences between the $R$-matrices
may be regarded as having their origins in different gradations
of the same algebra $U_q(A_1^{(1)})$.
The $R$-matrices (\ref{1/2,1/2-h}) and (\ref{1/2,1/2-p}) have
different limits when $u\to 0$, which can no longer be
transformed into each other by a similarity transformation, and
the associated braid group generators are inequivalent \cite{Wadati}\cite{BP}.

\sect{Infinitely Many Gauge-Equivalent $R$-Matrices for $U_q(A_2)$}
We now turn to  $U_q(A_2^{(1)})$ . In this case,
for any $z\in {\bf C}^\times$, there exist algebra homomorphisms
${\rm ev}_z$: $U_q(A_2^{(1)})\rightarrow U_q(A_2)$ given by
\begin{eqnarray}
&&{\rm ev}_z(E_\alpha)=z^{s_1}E_\alpha\,,~~{\rm ev}_z(F_\alpha)=
  z^{-s_1}F_\alpha\,,~~
  {\rm ev}_z(h_\alpha)=h_\alpha\nonumber\\
&&{\rm ev}_z(E_\beta)=z^{s_2}E_\beta\,,~~{\rm ev}_z(F_\beta)=
  z^{-s_2}F_\beta\,,~~
  {\rm ev}_z(h_\beta)=h_\beta\nonumber\\
&&{\rm ev}_z(E_{\delta-\alpha-\beta})=z^{s_0}
  F_{\alpha+\beta}q^{(h_\beta-h_\alpha)/3}
  \,,~~{\rm ev}_z(F_{\delta-\alpha-\beta})=z^{-s_0}q^{(h_\alpha-h_\beta)/3}
  E_{\alpha+\beta}\nonumber\\
&&{\rm ev}_z(h_{\delta-\alpha-\beta})=-h_{\alpha+\beta}\,,
  ~~{\rm ev}_z(c)=0\,,\label{g2}
\end{eqnarray}
where $s_0,~s_1$ and $s_2$ are arbitrary real numbers and define the
gradation of $U_q(A_2^{(1)})$.

Carrying out long but similar calculations to those given previously
\cite{ZG2},
we derive from (\ref{aR}) an infinite family of
$R$-matrices
associated with the fundamental representation of $U_q(A_2)$, corresponding
to the infinitely many different gradations $(s_0,s_1,s_2)$:
\begin{eqnarray}
R^{\rm (s_0,s_1,s_2)}_{(3),(3)}(u)&=&g_q(u)\cdot \left \{e_{11}+e_{55}+e_{99}
  +\frac{q^{-1}(1-u^{s_1+s_2+s_0})}{1-q^{-2}u^{s_1+s_2+s_0}}
  (e_{22}+e_{33}+e_{44}+e_{66}+\right .\nonumber\\
& &~~~~~~~~~~+e_{77}+e_{88})+\frac{1-q^{-2}}{1-q^{-2}u^{s_1+s_2+s_0}}
  (u^{s_1}e_{24}+u^{s_1+s_2}e_{37}+u^{s_2}e_{68}+\nonumber\\
& &~~~~~~~~~~\left .+u^{s_2+s_0}e_{42}+u^{s_0}e_{73}
  +u^{s_1+s_0}e_{86})\right \}\,,\label{33}
\end{eqnarray}
$e_{ij}$ is the matrix satifying $(e_{ij})_{kl}=\delta_{ik}
\delta_{jl}$ and
\begin{equation}
g_q(u)=q^{2/3}\cdot {\rm exp}\left (\sum_{n>0}\frac{q^{2n}-q^{-2n}}{q^{2n}+1
 +q^{-2n}}\,\frac{u^{(s_1+s_2+s_0)n}}{n}\right )
\end{equation}
is an irrelevant scalar factor which, like the other scalar factors, will be
ignored in what follows.

These various $R$-matrices  all satisfy the parameter-dependent QYBE,
and are apparently all associated with 15-vertex
models, solvable in principle. The following remarks are in order:
\\(i) Semi-classical limit:
\begin{eqnarray}
R^{(s_0,s_1,s_2)}_{(3),(3)}(u)&=&I+\left (\frac{1}{2}\log q\right )
  r^{(s_0,s_1,s_1)}_{(3),(3)}(u)+{\cal O}\left ((\frac{1}{2}\log q)^2
  \right )\,,\nonumber\\
r^{\rm (s_0,s_1,s_2)}_{(3),(3)}(u)&=&\frac{1+u^{s_1+s_2+s_0}}
  {1-u^{s_1+s_2+s_0}}(e_{11}+e_{55}+e_{99}-
  e_{22}-e_{33}-e_{44}-e_{66}-e_{77}-e_{88})\nonumber\\
&  &+\frac{4}{1-u^{s_1+s_2+s_0}}
  (u^{s_1}e_{24}+u^{s_1+s_2}e_{37}+u^{s_2}e_{68}+\nonumber\\
 &&~~~~~~~~ +u^{s_2+s_0}e_{42}+u^{s_0}e_{73}
  +u^{s_1+s_0}e_{86})\,.\label{classical2}
\end{eqnarray}
This defines the corresponding rational solutions of the CYBE.
\\(ii) Homogeneous gradation $s_1=s_2=0\,,~s_0=1$:  This produces the
$R$-matrix
\begin{eqnarray}
R^{\rm h}_{(3),(3)}(u)&=&e_{11}+e_{55}+e_{99}
  +\frac{q^{-1}(1-u)}{1-q^{-2}u}
  (e_{22}+e_{33}+e_{44}
  +e_{66}+e_{77}+e_{88})+\nonumber\\
& &~~~~~~~~~~+\frac{1-q^{-2}}{1-q^{-2}u}\left (e_{24}+e_{37}+e_{68}+
  u(e_{42}+e_{73}+e_{86})\right )\label{33-h}
\,,\end{eqnarray}
which is exactly the well-known solution due to Jimbo \cite{Jimbo}.
\\
(iii) Principal gradation $s_1=s_2=s_0=1$: This defines the
$R$-matrix
\begin{eqnarray}
R^{\rm p}_{(3),(3)}(u)&=&e_{11}+e_{55}+e_{99}
  +\frac{q^{-1}(1-u^3)}{1-q^{-2}u^3}
  (e_{22}+e_{33}+e_{44}
  +e_{66}+e_{77}+e_{88})+\nonumber\\
& &~~~~~~~~~~+\frac{u(1-q^{-2})}{1-q^{-2}u^3}\left (e_{24}+e_{68}+e_{73}
  +u(e_{37}+e_{42}+e_{86})\right )\label{33-p}
\,.\end{eqnarray}
(iv) Our results suggest that, associated with the fundamental representation
of $U_q(A_2)$, there is no fully symmetric $R$-matrix.
However, there is an  ``almost" symmetric $R$-matrix
corresponding to the gradation
$g3=(0,1,1)$,
\begin{eqnarray}
R^{\rm g3}_{(3),(3)}(u)&=&e_{11}+e_{55}+e_{99}
  +\frac{q^{-1}(1-u^2)}{1-q^{-2}u^2}(e_{22}+e_{33}+e_{44}+e_{66}+e_{77}+e_{88})
  +\nonumber\\
& &+\frac{u(1-q^{-2})}{1-q^{-2}u^2}(e_{24}+e_{37}+e_{42}+e_{73})
  +\frac{1-q^{-2}}{1-q^{-2}u^2}(e_{68}+u^2e_{86})\,.\label{33-g3}
\end{eqnarray}
(v) Of all the $R$-matrices (20), only $R^{\rm h}_{(3),(3)}(u)$ can be
{\em directly} derived by solving the Jimbo equations \cite{Jimbo} or by using
the Yang-Baxterization method \cite{Jones}\cite{ZGB}\cite{Ma et al} since
others do not have the intertwining property for the usual two coproducts
of $U_q(A_2)$.\\
(vi) The matrices $R^{\rm (s_0,s_1,s_2)}_{(3),(3)}(u)$ and
$R^{\rm h}_{(3),(3)}(u)$
can be transformed into each other by similarity (gauge) transformations,
\begin{eqnarray}
&&S(u)R^{\rm (s_0,s_1,s_2)}_{(3),(3)}(u)S^{-1}(u)=R^{\rm h}_{(3),(3)}
  (u^{s_1+s_2+s_0})\,,\nonumber\\
&&S(u)=e_{11}+e_{55}+e_{99}+u^{-s_1/2}e_{22}
  +u^{s_1/2}e_{44}\nonumber\\
&&~~~~~~~+u^{-(s_1+s_2)/2}e_{33}+u^{(s_1+s_2)/2}e_{77}+
  u^{-s_2/2}e_{66}+u^{s_2/2}e_{88}\,,\label{trans2}
\end{eqnarray}
Eq.(\ref{trans2}) defines a gauge symmetry of the QYBE. This implies the
gauge transformations have their origins in different gradations of the
same algebra.
\\(vii) When the spectral parameter goes to zero, we obtain
\begin{eqnarray*}
R^{\rm h}_{(3),(3)}(u=0)&=&e_{11}+e_{55}
  +e_{99}+q^{-1}(e_{22}+e_{33}+e_{44}+e_{66}
  +e_{77}+e_{88})\nonumber\\
& &  +(1-q^{-2})(e_{24}+e_{37}+e_{68})\nonumber\\
R^{\rm p}_{(3),(3)}(u=0)&=&e_{11}+e_{55}
  +e_{99}+q^{-1}(e_{22}+e_{33}+e_{44}+e_{66}
  +e_{77}+e_{88})\nonumber\\
R^{\rm g3}_{(3),(3)}(u=0)&=&e_{11}+e_{55}
  +e_{99}+q^{-1}(e_{22}+e_{33}+e_{44}+e_{66}
  +e_{77}+e_{88})+(1-q^{-2})e_{68}
\end{eqnarray*}
These limiting forms are not related by similarity transformations and define
three inequivalent braid group generators \cite{BP}. (In particular,
they have differing numbers of linearly independent
eigenvectors corresponding to the eigenvalue $q^{-1}$.)
While the first of these is the universal $R$-matrix for $U_q(A_2)$ in the
fundamental representation, the relationship of the second and third ones to
this quantum algebra is unclear.

\sect{Changing Gradations: General Case}
We have seen that quantum $R$-matrices with different spectral parameter
dependence can be obtained from the universal $R$-matrix of the associated
quantum affine algebra by choosing different gradations. The change from one
gradation to another can be achieved through spectral parameter dependent
similarity transformations, examples of which we have seen in (\ref{trans1})
and (\ref{trans2}).

One strong point of our method is that we obtain the spectral dependent
$R$-matrices in a universal form (i.e., as an element of $U_q({\cal G})
\otimes U_q({\cal G})$ and thus representation independent, see eq.
(\ref{uni1})). We can also write the transformation between different
gradations
in a universal form as follows.

Associated with a given gradation of $U_q(\hat{\cal G})$ there is an algebra
homomorphism ${\rm D}^{(s)}_z$:~
$U_q(\hat{\cal G})\rightarrow U_q(\hat{\cal G})\otimes {\bf C}(z,z^{-1})$
defined by
\begin{equation}
{\rm D}^{(s)}_z(E_i)=z^{s_i} E_i\,,~~~{\rm D}^{(s)}_z(F_i)=z^{-s_i}F_i\,,
  ~~~{\rm D}^{(s)}_z(h_i)=h_i\,,~~~i=0,\cdots,r\label{general-g1}
\end{equation}
We call $z$ the spectral parameter.The $s_i$ are arbitrary real numbers.
The homogeneous gradation is given by $s_i=\delta_{i0}$ and below the
homomorphism corresponding to this gradation will be denoted as
${\rm D}^{\rm h}_z$.

We define the operator
\begin{equation}
T^{(s)}(z)=z^{\chi^{(s)}}\,,~~~~~\chi^{(s)}\in {\cal H}_0\label{general-t}
\end{equation}
where ${\cal H}_0$ is the subspace of the Cartan subalgebra generated by
the $h_i\,,~i=1,2,\cdots, r$, i.e. without $h_0$. This operator transforms
the homogeneous gradation as follows
\begin{eqnarray}
&&T^{(s)}(z)\,{\rm D}^{\rm h}_z(E_i)\,(T^{(s)}(z))^{-1}=z^{(\chi^{(s)},
  \alpha_i)+\delta_{i0}}E_i\,,\nonumber\\
&&T^{(s)}(z)\,{\rm D}^{\rm h}_z(F_i)\,(T^{(s)}(z))^{-1}=z^{-(\chi^{(s)},
  \alpha_i)-\delta_{i0}}F_i\,,\nonumber\\
&&T^{(s)}(z)\,{\rm D}^{\rm h}_z(h_i)\,(T^{(s)}(z))^{-1}=h_i\,,~~~i=0,1,\cdots,r
\end{eqnarray}
We observe that this can be rewritten as
\begin{equation}\label{changegradation}
T^{(s)}(z)\,{\rm D}^{\rm h}_z(a)\,(T^{(s)}(z))^{-1}={\rm D}^{(s)}_{z'}(a)\,,
{}~~\forall a\in U_q(\hat{\cal G})
\end{equation}
with $z'=z^{1/\mu}$, provided that the $s_i$ are related to $\mu$ and
$\chi^{(s)}$ as follows:
\begin{eqnarray}
&&s_i=\mu(\chi^{(s)},\alpha_i)\,,~~~i=1,2,\cdots,r\nonumber\\
&&s_0=\mu(1+(\chi^{(s)},\alpha_0))\label{general-s}
\end{eqnarray}
Thus, by solving these equations for $\mu$ and $\chi^{(s)}$ we find from
(\ref{general-t}) the operator $T^{(s)}(z)$ which relates the homogeneous
gradation to an arbitrary gradation according to (\ref{changegradation}).
Note that such a change of gradation is accompanied by a change of
spectral parameter from $z$ to $z^{1/\mu}$.
To determine $\mu$, we take a linear combination of equations
(\ref{general-s}) and the equation $\sum^r_{i=0}\,n_i\alpha_i=0$, where
$n_i$ are the Kac-labels (\cite{Kac}, p.54), and find
$\mu=\sum_{i=0}^r\,n_i\,s_i$.

The spectral parameter dependent universal R-matrices $R^{(s)}(u)$
can be obtained
from the universal R-matrix $R$ of $U_q(\hat{\cal G})$ as
\begin{equation}
R^{(s)}(u)=({\rm D}_u^{(s)}\otimes I)R.
\end{equation}
They are related to the spectral parameter dependent R-matrix in the
homogeneous
gradation by
\begin{equation}
R^{\rm h}(u^\mu)=\left(T^{(s)}(u^\mu)\otimes 1\right)^{-1}R^{(s)}(u)
        \left(T^{(s)}(u^\mu)\otimes 1\right).
\end{equation}
Note that the universal spectral parameter dependent R-matrices
$R^{(s)}(u)$ are elements of $U_q(\hat{\cal G})\otimes U_q(\hat{\cal G})$.
It is only after the specialization to some finite dimensional
representations $\pi^\lambda$,~ $\pi^{\lambda'}$ of $U_q(\hat{\cal G})$
\begin{equation}
R_{\lambda\lambda'}^{(s)}(u)=(\pi_\lambda\otimes \pi_{\lambda'})R^{(s)}(u)
\end{equation}
that $R_{\lambda\lambda'}^{(s)}(u)$ can be viewed as a spectral dependent
R-matrix of the quantum algebra $U_q(\cal G)$. This is so
because the $U_q(\hat{\cal G})$ modules $V(\lambda)$ and $V(\lambda')$ are
automatically also (possibly reducible) modules of $U_q(\cal G)$.
However only some $U_q(\cal G)$-modules are also $U_q(\hat{\cal G})$-modules.
We call these modules "affinizable". Spectral dependent $U_q(\cal G)$
R-matrices exist only for affinizable modules. For an investigation
of affinizable modules see \cite{DZ94}.

It is easily checked that eq. (\ref{trans1}) and eq. (\ref{trans2}) can be
obtained from this by specializing to the particular representations.

\sect{Gradations in Quantum Affine Toda Theories}
Even though the spectral parameter dependent $R$-matrices for different
gradations are related by similarity transformations, they are not necessarily
equivalent physically. A nice example of this is furnished by the quantum
affine Toda field theories.

It is well-known that associated to every $\hat{\cal G}$ there is a 1+1
dimensional affine Toda field theory \cite{OT}, denoted $T(\hat{\cal G})$.
It is described by the field equations
\begin{equation}
\Box\phi=\frac{\sqrt{-1}}{\beta}\sum^r_{i=0}n_i\alpha_i\,e^{\sqrt{-1}\beta\,
  \alpha_i\cdot \phi}\label{toda-e}
\end{equation}
The field $\phi(x,t)$ takes values in ${\cal H}_0$, the subspace of the
Cartan subalgebra generated by the $h_i$, $i=1,\cdots,r$, i.e. without $h_0$.
$\beta$ is the coupling constant
and the $\alpha_i$ the simple roots.
For $\hat{\cal G}=A_1^{(1)}$ eq.(\ref{toda-e}) specializes
to the sine-Gordon (or affine Liouville) equation. The field equations
(\ref{toda-e}) have soliton solutions.

The affine Toda theory $T(\hat{\cal G})$ posesses symmetry generators
$E_i\,,\,F_i\,,\,h_i\,,~i=0,1,\cdots, r$, which generate the quantum affine
algebra $U_q(\breve{\cal G})$
\cite{BLe}. Here $\breve{\cal G}$ is the dual Lie algebra to $\hat{\cal G}$,
i.e., it is obtained by interchanging the roles of the roots and the  coroots.
The deformation parameter $q$ is determined by the coupling constant as
$q=e^{-\sqrt{-1}\pi/{\beta^2}}$. The central charge is zero.

We will now explain how the physically
relevant gradation is determined by Lorentz invariance. In quantum
theory each soliton solution of (\ref{toda-e}) gives rise to a one-particle
state $|a,\theta>$ in the Hilbert space, where $a$ labels the particle
and $\theta$ is the rapidity \footnote {The rapidity is related to the
two-momentum by $p_0=m\,ch(\theta)$ and $p_1=m\,sh(\theta)$,~$m$ being
the mass of the particle.}. The defining property of a particle is its
behaviour under Lorentz transformations, which in two dimensions takes
the form,
\begin{equation}\label{lorentzonstate}
L(\lambda)|a,\theta>=|a,\theta+\lambda>
\end{equation}
where $L(\lambda)$ is the Lorentz generator. Also the transformation property
of the symmetry generators can be determined and one finds \cite{BLe}
\begin{equation}\label{lorentzoncharge}
L(\lambda) E_i=e^{\lambda s_i}E_i L(\lambda)\,,~~~
L(\lambda) F_i=e^{-\lambda s_i}F_i L(\lambda)\,,~~~
L(\lambda) h_i=h_i L(\lambda)\,,
\end{equation}
with
\begin{equation}
s_i=\frac{2}{(\alpha_i,\alpha_i)}\frac{1}{\beta^2}-1\label{general-s2}
\end{equation}
Comparing (\ref{lorentzoncharge}) and (\ref{lorentzonstate})
fixes the $\theta$-dependence of the action of the $U_q(\breve{\cal G})$
generators on the soliton states
\begin{eqnarray}
&&E_i|a,\theta>=e^{s_i\theta}\,\Pi(E_i)^a_{b}|b,\theta>\nonumber\\
&&F_i|a,\theta>=e^{-s_i\theta}\,\Pi(F_i)^a_{b}|b,\theta>\nonumber\\
&&h_i|a,\theta>=\Pi(h_i)^a_{b}|b,\theta>\label{reps}
\end{eqnarray}
where $\Pi$ is a $\theta$-independent finite dimensional representation of
$U_q(\breve{\cal G})$. We recognize the representation (\ref{reps}) as the loop
representation of $U_q(\breve{\cal G})$ with spectral parameter $e^\theta$ and
gradation $s_i$ given by (\ref{general-s2}). When $\hat{\cal G}$ is simply
laced, this is just the principal gradation (up to a rescaling of the
spectral parameter), but for non-simply laced theories, this is an
unusual gradation.

The physical quantity which is most immediately determined by the quantum
affine algebra symmetry is the scattering matrix which describes the
transition from an incoming 2-soliton state to an outgoing 2-soliton
state. As is explained in \cite{BLe}, this is proportional to the
$U_q(\breve{\cal G})$ $R$-matrix in the representation (\ref{reps}).
To predict the correct scattering behaviour of the solitons it is thus
essential to work with the $R$-matrix in the gradation determined by
(\ref{general-s2}). Different gradations are {\em not} physically
equivalent because physics singles out a particular basis in Hilbert
space, namely that given by particle states.
The non-standard gradation (\ref{general-s2}), taken together with the
axioms of $S$-matrix theory such as crossing symmetry, leads to
interesting effects in the non-simply laced case, as we will describe
in \cite{gustav}. In particular it determines the quantum mass-ratios
of the solitons.

\sect{Concluding Remarks}
We have found infinitely many spectral-dependent $R$-matrices
corresponding to different gradations --
including the important
homogeneous gradation and principal gradation -- of $U_q(A_1^{(1)})$
and $U_q(A_2^{(1)})$. These $R$-matrices are related to each other by
similarity (gauge) transformations but can
have quite different limits as the spectral parameter $u\rightarrow 0$,
and the choice of gradation can be dictated by the physics in
particular applications.

Our results suggest that the ``hierarchies" of solutions of the QYBE, which are
gauge equivalent for non-zero values of the spectral parameter,
but which may be inequivalent in the limit $u\to 0$, have their origin in
the gradations of the quantum affine algebras.

\vskip.3in
\begin{center}
{\bf Acknowledgements:}
\end{center}
We are grateful to Uwe Grimm, Valeriy N. Tolstoy and S.O.Warnaar for reading
the manuscript and pointing out some errors in the original version, and to
Valeriy N. Tolstoy for clarifying discussions. G.W.D. wishes
to thank the Department of Mathematics of the University of Queensland in
Brisbane for its hospitality. Y.Z.Z. wishes to thank Loriano Bonora,
G\"unter von Gehlen and Zhong-Qi Ma for discussions.
Financial support from the Australian Research Council is gratefully
acknowledged.

\vskip.3in


\begin{thebibliography}{99}
\bibitem{Drinfeld} V.G.Drinfeld, {\em Proc. ICM, Berkeley} {\bf 1} (1986) 798
\bibitem{Jimbo} M.Jimbo, {\em Lett.Math.Phys.} {\bf 10} (1985) 63,
  {\em ibid} {\bf 11} (1986) 247;
  {\em Commun.Math.Phys.} {\bf 102} (1986) 537
\bibitem{Reshetikhin} N.Reshetikhin, {\em Quantized universal enveloping
  algebras, the Yang-Baxter equation and inveriants of links: I, II},
  preprints LOMI E-4-87, E-17-87
\bibitem{Jones} V.F.R.Jones, {\em Int.J.Mod.Phys.} {\bf B4} (1990) 701
\bibitem{ZGB} A.J.Bracken, M.D.Gould and R.B.Zhang, {\em Mod.Phys.Lett}
  {\bf A5} (1990) 831; R.B.Zhang, M.D.Gould and A.J.Bracken, {\em Nucl.Phys.}
  {\bf B354} (1991) 625
\bibitem{Ma et al} Z.-Q.Ma, {\em Yang-Baxter equation and quantum enveloping
  algebras}, World Scientific, Singapore, 1993 and references therein;
  Z.-Q.Ma and A.-Y.Dai, {\em A new solution of the YBE related to the adjoint
  representation of $U_qB_2$}, Beijing IHEP preprint, 1993
\bibitem{ZG2} Y.-Z.Zhang and M.D.Gould, {\em Quantum affine algebras and
  universal $R$-Matrix with spectral parameter},
  preprint hep-th/9307007, {\em Lett.Math.Phys.} (in press)
\bibitem{Belavin} A.A.Belavin and V.G.Drinfeld, {\em Funct.Anal.Appl.}
  {\bf 16} (1982) 159
\bibitem{KT} S.M.Khoroshkin and U.N.Tolstoy, {\em Lett.Math.Phys.}
  {\bf 24} (1992) 231; {\em Funkz.Analyz. i ego Pril.} {\bf 26} (1992) 85
\bibitem{ZG} Y.-Z.Zhang and M.D.Gould,
  {\em Lett.Math.Phys.} {\bf 29} (1993) 19
\bibitem{Wadati} K.Sogo, M.Uchinami, Y.Akutsu and M.Wadati,
  {\em Prog.Theor.Phys.} {\bf 68} (1982) 508;
  M.Wadati, T.Deguchi and Y.Akutsu, {\em Phys.Rep.}
  {\bf 180} (1989) 247
\bibitem{BP} V.Bazhanov, {\em communications}; P.A.Pearce, {\em communications}
\bibitem{Kac} V.Kac, `Infinite dimensional Lie algebras', third edition,
  Cambridge University Press (1990).
\bibitem{DZ94} G.W.Delius and Y.-Z.Zhang, {\em Finite dimensional
  representations of quantum affine algebras}, preprint hep-th/9403162
\bibitem{OT} D.Olive and N.Turok, {\em Nucl.Phys.} {\bf B220} (1983) 491
  and {\bf B257} (1985) 277; H.W.Braden, E.Corrigan, P.E.Dorey and R.Sasaki,
  {\em Phys.Lett.} {\bf B227} (1989) 441
\bibitem{BLe} D.Bernard and A.LeClair, {\em Commun.Math.Phys.} {\bf 142}
  (1991) 99
\bibitem{gustav} G.W.Delius, in preparation
\end{thebibliography}
\end{document}